\documentstyle[11pt]{article}
\newcommand{\blankline}{\vskip .3cm}
\newcommand{\f}{\begin{equation}}
\newcommand{\ff}{\end{equation}}
\begin{document}
\rightline{hep-th/9411064}
\rightline{CGPG-94/9-3, VPI-IHEP-94/8}
\blankline
\centerline{\LARGE Chiral Fermions, Gravity and GUTs\rm${}^@$}
\blankline
\rm
\vskip.3cm
\centerline{Lay Nam Chang${}^*$ and Chopin Soo ${}^\dagger$}
\blankline
\centerline{${}^*$Institute for High Energy Physics and Dept. of Physics}
\centerline{Virginia Polytechnic Institute and State University}
\centerline{Blacksburg, VA 24061-0435}
\vskip.3cm
\centerline{${}^\dagger$Center for Gravitational Physics and Geometry}
\centerline{Department of Physics}
 \centerline {Pennsylvania State University}
\centerline{University Park, PA 16802-6300}
 \blankline
\blankline
\blankline
\blankline
\centerline{\bf ABSTRACT}
\noindent
\blankline

We discuss a global anomaly associated with
the coupling of chiral Weyl fermions to gravity.
The Standard Model based upon
$SU(3){\times}SU(2){\times}{U(1)}$ which has 15 fermions
per generation is shown to be inconsistent if all background spin
manifolds with signature invariant $\tau=8k$ are allowed. Similarly,
GUTs based on odd number of fermions are inconsistent. Consistency can
be achieved by adding an extra Weyl fermion which needs to couple only
to gravity. For arbitrary $\tau$'s, generalized
spin structures are needed, and the global anomaly cancellation requires
that the net index of the total Dirac operator with spin and internal
gauge connections be even. As a result GUTs with fundamental multiplets
which contain multiples of 16 Weyls per generation are selected. The
simplest consistent GUT is the SO(10) model with a multiplet of 16 Weyls per
generation. The combined gravity and internal symmetry gauge group of the
theory is then $[Spin(3,1){\times}Spin(10)]/Z_2$. Physical implications of
these results are commented on.

\vfill
${}^@$To appear in the {\it Proceedings of the 4th. Drexel Symposium 1994}
\blankline
${}^*$laynam@lotus.cc.vt.edu; $\quad{}^\dagger$ soo@phys.psu.edu
\eject

Witten\cite{Witten} uncovered a global anomaly which states that
in 4 dimensions, a gauge theory with an odd number of chiral fermion
doublets coupled to $SU(2)$ gauge fields is inconsistent.
In the Standard Model based
upon ${SU_C(3)}{\times}{SU_W(2)}{\times}{U_Y(1)}$,
the number of left-handed fermion doublets coupled to $SU_W(2)$
turns out to be reassuringly even (4 per generation). However,
when the gravitational field is taken into account,
there is an {\it additional} $SU(2)$ gauge group
which comes from the rotation subgroup of the local
Lorentz gauge group of gravity. An immediate question is if
additional constraints need to be imposed on models of Particle Physics when
interactions with the gravitational fields are included\cite{cs}.
In particular,
these consistency conditions can arise from treating Weyl fermions as quantum
fields in gravitational backgrounds of arbitrary topologies.
Witten's arguments have been elaborated on by others\cite{deAlwis}.
We review the essential points, and show how an
inconsistency can arise. The relevant doublets for the case we
have at hand are the two components of Weyl fermions.

To begin with, consider a suitable Wick rotation of the
background spacetime into a Riemannian manifold, or the more general
setting of matter coupled to gravity in Euclidean Quantum Gravity. The
fermions can be expanded in terms of the complete set
of the eigenfunctions,$\{X_n\}$, of the hermitean Dirac operator
\cite{Fujikawa}. The generating functional,
$Z_W[\omega]$,
for a left-handed 2-component Weyl doublet
coupled to the gravitational spin connection, ${\omega}$,
can be defined to be the square-root of the generating functional of a
bispinor. (More precisely, one can think of each Weyl doublet coupled
to gravity as a Majorana bispinor). We then have
\f
Z_W[\omega] = Z^{1\over 2}_{Dirac}[\omega]
= \{det(i\gamma^\mu D^\omega_{\mu})\}^{1\over 2}
\ff

Consider next a chiral transformation by $\pi$ which maps each 2-component
left-handed Weyl fermion $\Psi_L(x)\mapsto\ exp(i\pi\gamma_5)\Psi_L(x) =
\exp(-i\pi)\Psi_L(x)= -\Psi_L(x)$. Obviously such a map is a symmetry of
the action. However, the measure is not necessarily invariant
\cite{Fujikawa} under such a chiral transformation because of the
Adler-Bell-Jackiw anomaly\cite{ABJ}. Instead, each left-handed fermion
 measure transforms as
\f
d\mu \mapsto d\mu \exp(-i\pi \int_M d^4x \det(e)
\sum_n X^\dagger_n(x)\gamma_5X_n(x))
\ff
The expression
$\int_M d^4x \sum_n \det(e) X^\dagger_n(x)\gamma_5X_n(x)$,
(which needs to be regularized) is
formally equal to $(n_+ - n_-)$, where $n_{\pm}$ are the number of
normalizable positive and negative chirality zero
modes of the Dirac operator.
Upon regularization, the expression works out to be\cite{Fujikawa}
\begin{eqnarray}
\sum_n {\det(e)}X^\dagger_n{\gamma_5}{X_n}&\equiv&
\lim_{{\cal M} \rightarrow \infty}
\sum_n \det(e) X^\dagger_n(x){\gamma_5}e^{-(\lambda_n/ {\cal M})^2}X_n(x)
\nonumber\\
&=&\lim_{{\cal M} \rightarrow \infty}\lim_{x' \rightarrow x}Tr\gamma_5
\det(e)e^{-(i\gamma^\mu{D_\mu}/{\cal M})^2}\sum_n X_n(x)X^\dagger_n(x')
\nonumber\\
&=& -{1\over 384\pi^2}R_{\mu\nu\sigma\tau}\ast R^{\mu
\nu\sigma\tau}
\end{eqnarray}
As a result,
$(n_+ - n_-) = \sum_n \int_M d^4x \det(e) X^\dagger_n(x)\gamma_5X_n(x) =
-\tau/8$, where $\tau = (1/48\pi^2)\int_M R_{\mu\nu\sigma\tau}\ast
R^{\mu \nu\sigma\tau}$, is the signature invariant of the four-manifold $M$.
(In this discussion, we shall restrict our attention to compact orientable
4-manifolds without boundary.)
If there are altogether $N$ left-handed Weyl fermions in
the theory, the total measure changes by
$\exp(iN \pi\tau/8)$.

But, the chiral transformation of $-1$ on the fermions can also be
considered to be an ordinary $SU(2)$ rotation of $2\pi$.
Since $SU(2)$ is a safe group with no perturbative chiral anomalies
(there are no Lorentz anomalies in 4 dimensions), the measure must be
invariant under all $SU(2)$ transformations. Thus an {\it inconsistency}
arises unless the phase factor, $\exp(iN \pi\tau/8)$, is always unity.

It is known that for consistency of parallel
transport of spinors for topological four-manifolds, $\tau$ must be a
multiple of 8 for spin structures to exist. If in quantum
gravity, or in semiclassical quantum field theories, one allows only for
four-manifolds with $\tau = 16k, k \in Z$; then under a chiral
rotation of $\pi$ the measure is invariant regardless of $N$. Otherwise,
consistency with the global anomaly requires that $N$ must be even if all
spin manifolds with $\tau =8k$ are permitted.
If one counts the number of left-handed Weyl fermions in the Standard Model,
one finds that the number is 15 per generation, giving a
total of 45 for 3 generations.
This comes about because each bispinor is coupled twice to the spin
connection while each Weyl spinor is coupled once (e.g. for the first
generation, the number is 2 for each electron and each up or down quark of a
particular color, and 1 for each left-handed neutrino.)
Thus even if one restricts to $\tau =8k$, the global anomaly with respect
to $SU(2)$ rotations implies that there should be additional
particle(s). For example, there could be a partner
for each neutrino, making $N$ to be 16 per generation, or a partner for
just the $\tau$-neutrino, or even four generations.
As a result, even with $\tau = 8k$, grand unification schemes based upon
groups such as SU(5) and odd number of Weyl fermions would be inconsistent
when coupled to gravity.

It is quite likely that in quantum gravity the
allowed manifolds should extend over a wider class than those which admit
classical spin structures\cite{Hawking} .
The considerations outlined above suggest that in order to do this,
we would have to allow for couplings via gauge fields among the various
fermions, which is what occurs in Grand Unified
Theories (GUTs). The reason can be stated as follows.
When $\tau$ is not a multiple of 8, it is not possible to lift the
$SO(4)$ bundle to its double cover $Spin(4)$ bundle since the second
Stiefel-Whitney class is non-trivial. However, given a general grand
unification simply-connected gauge group $G$ with a $Z_2 =
\{e, c\}$ in its center, it is possible to construct
{\it generalized spin structures} with gauge group $Spin_G(4) =
{\{Spin(4)\times G\}/{Z_2}}$ where the $Z_2$ equivalence relation
is defined by $(x,g) \equiv (-x,cg)$ for
all $(x,g) \in Spin(4)\times G$\cite{Hawking}. Note that $Spin_G(4)$ is the
double cover of $SO(4)\times (G/Z_2)$. The parallel transport of fermions
then does not give rise to any inconsistency.

Now, in a scenario where all of the fermions
are coupled to one another,
the index theorem should be applied only to the trace current
involving the sum over all the fermion fields. The resultant restriction on
$N$ is the condition that the index for the {\it total}
Dirac operator with $Spin_G(4)$ connections, $ N_+ - N_-$, is even.
The index is given by
\f
N_+ - N_- = -N\tau/8 - {1/({8\pi^2})}\int_M Tr (F\wedge F)
\ff
where $F$ is the curvature of the GUT group and $N$ is the total number of
Weyl fermions in the GUT multiplet.
The net conclusion is therefore that inclusion of manifolds with
arbitrary $\tau$'s is possible, provided we allow
for generalized spin structures, with a total of $16k$ fermions
and a GUT-group with always even instanton numbers,
in order to ensure that the global anomaly is absent.

In the event that the structures are defined by simple GUT groups, the
preeminent choice would be $SO(10)$\cite{Georgi}. It is easy to check that
the 16 Weyl fermions in
the $SO(10)= Spin(10)/{Z_2}$ GUT indeed belong to a 16-dim. representation
of $Spin(10)$, and satisfy the generalized spin structure
equivalence relation for
$\{Spin(4)\times Spin(10)\}/{Z_2}$. Also, ${1/({8\pi^2})}\int_M Tr
(F\wedge F)$ is always even.
It is worth emphasizing that the generalized spin structure implies an
additional ``isospin-spin'' relation in that
fermions must belong to $Spin(10)$
representations, while bosons must belong to $SO(10)$ representations of the
GUT. This has implications for spontaneous symmetry breaking via
fundamental bosonic Higgs, which cannot belong to the spinorial
representations of $SO(10)$ if one allows for arbitrary $\tau$'s.
More generally,  the presence of the extra particle(s) implied by
global anomaly considerations can generate masses for neutrinos,
thereby giving rise to neutrino
oscillations, and also play a significant role in the cosmological
issue of  ``dark matter".

We have presented our arguments in terms of conventional spin connection
couplings for definiteness. We would like
to end by emphasizing that our results would also obtain\cite{cs}
in the setting of Weyl fermions coupled to Ashtekar-Sen
connections in  the Ashtekar formulation of gravity \cite{ashtekar}.
\section*{Acknowledgments}
This work is supported in part by the Department of Energy under Grant
No. DE-FG05-92ER40709, the NSF under Grant No. PHY-9396246, and research
funds provided by the Pennsylvania State University.

\end{document}